\documentstyle[epsf,prd,aps,floats]{revtex}
\newcommand{\be}{\begin{equation}}
\newcommand{\ee}{\end{equation}}

\newcommand {\bea}{\begin{eqnarray}}
\newcommand {\eea}{\end{eqnarray}}
\def\Mpc{{\rm Mpc}}
\def\etal{{\frenchspacing\it et al.}}
\def\ie{{\frenchspacing\it i.e.}}

\def\etc{{\frenchspacing\it etc.}}

\def\beq#1{\begin{equation}\label{#1}}
\def\eeq{\end{equation}}
\def\beqa#1{\begin{eqnarray}\label{#1}}
\def\eeqa{\end{eqnarray}}
\def\eq#1{equation~(\ref{#1})}

\def\fig#1{Figure~\ref{#1}}

\def\spose#1{\hbox to 0pt{#1\hss}}
\def\simlt{\mathrel{\spose{\lower 3pt\hbox{$\mathchar"218$}}
     \raise 2.0pt\hbox{$\mathchar"13C$}}}
\def\simgt{\mathrel{\spose{\lower 3pt\hbox{$\mathchar"218$}}
     \raise 2.0pt\hbox{$\mathchar"13E$}}}
\def\simpropto{\mathrel{\spose{\lower 3pt\hbox{$\mathchar"218$}}
     \raise 2.0pt\hbox{$\propto$}}}

\def\Om{\Omega_m}
\def\Oms{\Om^*}
\def\rhos{\rho_*}
\def\ws{w_*}
\def\k{{\bf k}}

\begin{document}
\twocolumn[\hsize\textwidth\columnwidth\hsize\csname@twocolumnfalse\endcsname
%\title{Growth of inhomogeneities in a generalized Chaplygin gas Universe}
\title{The End of Unified Dark Matter?}
\author{H\aa vard B. Sandvik$^1$, Max Tegmark$^1$, Matias
Zaldarriaga$^2$ \& Ioav Waga$^3$}
\address{$^1$Dept. of Physics, Univ. of Pennsylvania, Philadelphia, PA
19104, USA;sandvik@hep.upenn.edu\\ 
$^2$Dept. of Physics, New York University, 4 Washington Pl., New York, NY 10003\\
$^3$Universidade Federal do Rio de Janeiro, Instituto de F\'\i sica,
Cep 21041-972, Brazil}

\maketitle

\begin{abstract}
Despite the interest in dark matter and dark energy,
it has never been shown that they are in fact two separate substances. 
We provide the first strong evidence that they are separate by
ruling out a broad class of so-called unified dark matter 
models that have attracted much recent interest.
%We show for the first time that dark matter and dark energy
%are separate substances, thereby ruling out a broad class of 
%so-called unified dark matter models that have attracted much recent interest.
We find that they produce oscillations or exponential blowup of the matter power spectrum
inconsistent with observation.
For the particular case of generalized Chaplygin gas models, 99.999\% of the
previously allowed parameter space is excluded, 
leaving essentially only the standard $\Lambda$CDM limit allowed. 
\end{abstract}
\vspace{2cm}
]

\section{Introduction}

Despite the broad interest in dark matter and dark energy, their physical properties
are still poorly understood. Indeed, it has never even 
been shown that the two are in fact two separate substances. 
The goal of this paper is to show that they are.

There is strong evidence from a multitude of observations that 
there is about six times more cold dark matter (CDM) than baryons in
the cosmic matter
budget, making up of order 30\% of critical density
\cite{consistent,efstathiou}. 
In addition to this clustering dark component, observations of 
supernovae, the cosmic microwave background fluctuations and galaxy clustering 
provide mounting evidence of 
a uniformly distributed dark energy with negative 
pressure which has come to dominate the universe 
recently (at redshifts $z\lesssim 1$) and caused its expansion to accelerate. 
It currently constitutes about two thirds of the critical density
\cite{consistent,efstathiou,RatraPeebles}.

Although the dark energy can be explained by introducing the
cosmological constant ($\Lambda$) into general relativity (lending the 
standard model the name $\Lambda$CDM),
this ``solution'' has two severe problems, frequently triggering 
anthropic explanations 
% Cite Weinberg; Efstathiou; Vilenkin 
and general unhappiness.
The first problem is explaining its magnitude,
since theoretical predictions for
$\Lambda$ lie many orders of magnitude above the observed value.
The second problem is the so-called
cosmic coincidence problem: explaining why the
three components of the universe (matter, radiation and $\Lambda$)
presently are of similar magnitudes although they all scale differently
with the Universe's expansion.

As a response to these problems, much interest has been devoted to
models with dynamical vacuum energy, so-called
quintessence\cite{quintessence}. These models typically involve
scalar fields with a particular class of potentials, allowing the
vacuum energy to become dominant only recently. Although
quintessence is the most studied candidate for the dark
energy, it generally does not avoid fine tuning in explaining the
cosmic coincidence problem. Recently several
alternative models have been proposed, such as the condensate
models of\cite{condensates}.

%poses the problem of explaining
%One of the main problems in modern cosmology is to explain the
%present acceleration of the Universe. There is 
%
%There is increasing evidence for a cosmological model with a
%Universe dominated at present by a smooth dark-energy component.
%The main theoretical attempts at explaining this effect have
%concentrated on models with dynamical vacuum energy, so-called
%quintessence\cite{quintessence}. These models typically involve
%scalar fields with a special class of potentials, allowing the
%vacuum energy to become dominant only recently. Although
%quintessence is the most exhausted explanation for the dark
%energy, it generally does not avoid the fine tuning problems in
%explaining the cosmic coincidence problem. Recently several
%alternative models have been proposed, such as the condensate
%models of\cite{bassett}.

An alternative to quintessence which has attracted great
interest lately is the so-called generalized Chaplygin
gas (hereafter GCG) 
\cite{kamenshchik,bento,fabris,bilic,gorini,dev} (see also the related
earlier work of\cite{barrow}).
Rather than fine tuning some potential, the model explains the
acceleration of the Universe via an exotic equation of
state causing it to act like dark matter at high density and 
like dark energy at low density.
The model is interesting for phenomenological reasons but can
be motivated by a brane-world interpretation\cite{bilic,bento}.
An attractive feature of the model is that it can
explain both dark energy and dark matter in terms of 
a single component, and has therefore been referred to as
unified dark matter (UDM) or ``quartessence'' \cite{makler}.
(See also \cite{padmanabhan}.)

This approach has been thoroughly investigated for its impact on the 0th order 
cosmology, i.e., the cosmic expansion history (quantified by the
Hubble parameter $H[z])$ and  
corresponding spacetime-geometric observables.
An interesting range of models was found to be consistent with 
SN Ia data \cite{makler} and CMB peak locations \cite{bento2}.

Some work has also studied constraints from 1st order cosmology
(the growth of linear perturbations), finding 
an interesting range of models to be consistent with cosmic microwave background
(CMB) measurements \cite{carturan}. There is, however, a fatal flaw 
in UDM models that manifests itself only at recent times and on smaller 
(Galactic) scales and has therefore not been revealed by these studies.
As we will see, this flaw rules out all GCG models except those
that are for all practical purposes identical to the 
usual $\Lambda$CDM model.

% PRL doesn't allow sections! (Silly, I know... :)
The rest of this {\it Letter} is organized as follows. 
In the next section, we review 
the fundamentals of the GCG model. We then consider in section
\ref{pert} the evolution of density inhomogeneities in the model
and use the predicted matter power spectrum to constrain it with
observational data. 
Finally, we describe how the basic flaw that rules out the 
GCG model is indeed a generic feature of a broad class of unified dark matter models.

\section{The Chaplygin gas}\label{chg}

%In this section we lay out the fundamentals of the model.
A standard assumption in cosmology is that the pressure of a single substance is,
at least in linear perturbation theory, uniquely determined by its density. 
A generalized Chaplygin gas 
%(hereafter GCG) 
\cite{kamenshchik,bilic,fabris} 
is simply a substance where this
relation $p(\rho)$ is a power law
\beq{eqnofstate}
p = - A \rho^{-\alpha}
\eeq
with $A$ a positive constant. 
The original Chaplygin gas had $\alpha=1$.
The standard $\Lambda$CDM model has two separate dark components, both with  $\alpha=-1$, giving a 
constant equation of state $w\equiv p/\rho$ that equals $0$ for dark matter and $-1$ for dark energy.

By inserting \eq{eqnofstate} into the energy conservation law, one finds that the GCG
density evolves as \cite{makler}
\beq{rhoEq}
\rho(t) = \left[A + \frac{B}{a^{3(1+\alpha)}}\right]^{1 \over 1+\alpha},
\eeq
where $a(t)$ is the cosmic scale factor normalized to unity today, i.e., 
$a=(1+z)^{-1}$ where $z$ denotes redshift.
Here $B$ is an integration constant. The striking feature here is that although
the GCG has $\rho\propto a^{-3}$ when sufficiently compressed, 
it's density will never drop below the value $A^{1/1+\alpha}$ 
no matter how much you expand it. 
%If we parametrize the
%equation of state in the usual way $\rho = w p$ we find for the
%time varying $w(a)$:
%\be
%w = - \frac{A}{\rho(a)^{1+\alpha}}
%\ee
Defining
\beq{RedefEq}
\Oms\equiv {B\over A+B},\quad \rhos\equiv (A+B)^{1\over 1+\alpha},
\eeq
\eq{rhoEq} takes the form
\beq{rhoEq2}
\rho(a) = \rho_*\left[(1-\Omega_m^*) + \Omega_m^* a^{-3(1+\alpha)}\right]^{1 \over 1+\alpha}.
\eeq
For comparison, a standard flat model with current CDM density parameter $\Om$
as well as dark energy density $(1-\Om)$ whose equation of state $w_*$ is constant gives
\beq{rhoEq3}
\rho(a) = \rho_*\left[(1-\Omega_m) a^{-3(1+\ws)} + \Omega_m a^{-3}\right].
\eeq
We see that the last two equations bear a striking similarity even though
the former involves a single substance and the latter involves two.
Both have two free parameters. Both have the current density $\rho(1)=\rho_*$.
Making the identification $\Oms=\Om$, both have
$\rho(a)\to\Om\rhos a^{-3}$ at early times as $a\to 0$ (for $\ws<0$),
showing that $\Oms$ can be interpreted as an effective matter density in the GCG model.
Indeed, for the special case $\alpha=0$ and $\ws=-1$, we see that both models coincide 
with standard $\Lambda$CDM. 
For $\alpha=0$ the GCG model becomes equivalent to $\Lambda$CDM 
not only to 0th order in perturbation theory as above but to all orders, 
even in the nonlinear clustering regime.

The 0th order cosmology determined by \eq{rhoEq2} together with the Friedman equation 
\beq{FriedmanEq}
H\equiv {\dot a\over a} = \left[{8\pi G\over 3}\rho\right]^{1/2}
\eeq
(which determines $a(t)$ and the spacetime metric to 0th order) 
has been thoroughly in previous
work\cite{kamenshchik,bilic}, 
and by studying constraints from supernovae observations, Makler
{\etal} \cite{makler} have placed interesting constraints on the
$(\alpha,\Oms)$ parameter space.

\section{Growth of inhomogeneities}\label{pert}

Let us now consider the evolution of density perturbations in this
UDM model. Following the standard calculations of
\cite{padmanabhan}, we obtain for the relativistic analog of the
Newtonian 1st order perturbation equation in Fourier space that
a density fluctuation $\delta_k$ with wave vector $\k$ evolves
as 
\beqa{GrowthEq}
&&\ddot{\delta}_k+H \dot{\delta}_k[2-3(2w-c_s^2)]\\ \nonumber
&&-\frac32H^2\delta_k\left[ 1-6c_s^2-3w^2+8w\right] = -\left({k c_s \over
a}\right)^2\delta_k,
\eeqa
where the equation of state $w\equiv p/\rho$ and 
the squared sound speed $c_s^2\equiv \partial p/\partial\rho$ are
evaluated to 0th order and hence depend only on time, not on position. 
(We use units where the speed of light $c=1$ throughout.)
This equation is valid on subhorizon scales $|\k|\gg H/c$.
In other words, the growth of density fluctuations is completely 
determined by the two functions $w(a)$ and $c_s^2(a)$.
Combining \eq{eqnofstate} and \eq{rhoEq2}, these two functions are
\cite{makler}
\beqa{wEq}
w    &=& -\left[1+{\Oms \over 1-\Oms}a^{-3(1+\alpha)}\right]^{-1},\\
c_s^2&=& -\alpha w = \alpha \left[1+{\Oms \over 1-\Oms}a^{-3(1+\alpha)}\right]^{-1}.
\eeqa
This shows a second reason why the GCG has been considered promising for cosmology
it starts out behaving like pressureless CDM 
($w\approx 0$, $c_s\approx 0$) early on (for $a\ll 1$)
and gradually approaches cosmological constant behavior ($w\approx -1$) at 
late times. There is also an intermediate state where the
effective equation of state is $p=\alpha\rho$\cite{kamenshchik}.
(Going beyond 1st order perturbation theory, the GCG that gets 
gravitationally bound in galactic halos maintains its density high enough
to keep acting like CDM forever.)

To solve \eq{GrowthEq} numerically, we change the independent variable from
$t$ to $\ln a$. Using the properties
\be
{d\over dt} = H {d\over d\ln a},
\quad
\ddot{\delta}_k = H^2 \delta'' + {1 \over 2}(H^2)' \delta',
\ee
where $'\equiv d/d\ln a$ and 
\be
\xi\equiv {(H^2)' \over 2H^2} = -\frac32\left(1+(1/\Oms-1)a^{3(1+\alpha)}\right)^{-1},
\ee
\eq{GrowthEq} takes the form
\bea
\delta_k''& + &[2+\xi -3(2w-c_s^2)]\delta_k'\\ \nonumber
&=& \left[\frac 32
(1-6c_s^2+8w-3w^2)-\left({k c_s\over aH}\right)^2
\right]\delta_k. \\ \label{firstorder}
\eea
Even before solving this, it s obvious that 
a non-zero sound speed, if
present for a sufficiently long time-span, is going to have a
dramatic effect on the $k$-dependence of the perturbation growth. 
If $c_s^2>0$, then fluctuations with wavelength below the
Jeans scale $\lambda_J = \sqrt{\pi |c_s^2|/G\rho}$ will be pressure-supported
and oscillate rather than grow. 
This oscillation is confirmed by the numerical solutions, and is analogous to
the acoustic oscillations in the photon-baryon fluid in the pre-decoupling epoch.
If $c_s^2<0$, corresponding to negative $\alpha$, fluctuations below this wavelength will be violently unstable
and grow exponentially \cite{HuGDM}.

A key point which has apparently been overlooked in prior work is that
whereas all the other terms in \eq{firstorder}
are of order unity or smaller, the sound speed term 
$(k c_s/aH)^2$ can be much larger even if the sound speed is
tiny $|c_s|\ll 1$.
This is because $c_s$ is multiplied by the prefactor $k/aH$ which can be enormous,
since it it the Horizon scale divided by the perturbation scale.
\def\lambdac{\lambda_c}
Defining a critical wavelength $\lambdac$ by
\beq{lambdacritDefEq}
\lambdac^2\equiv {c_s^2\over (aH)^2} = -{\alpha w\over (aH)^2},
\eeq
the pressure term
% $(k c_s/aH)^2$ 
in \eq{firstorder} becomes simply $(\lambdac k)^2$,
so we expect oscillations or exponential blowup in the power spectrum 
on scales $k\simgt \lambdac^{-1}$. These are created mainly during the 
recent transition period when both $a$ and $-w$ are or order unity (growing 
from 0 to 1), and since neither effect is seen in observed data,
we therefore expect to obtain constraints of order 
$|\alpha|\simlt (H/k)^2$, 
the squared ratio of the perturbation scale to the horizon scale.
This heuristic argument thus suggests that Galaxy clustering constraints on scales down to 
$10 h^{-1}\Mpc$ would give the constraint
$|\alpha|\simlt (10 h^{-1}\Mpc/3000 h^{-1}\Mpc)^2\approx 10^{-5}$ --- we will see that this
approximation is in fact fairly accurate.

\begin{figure}
\centerline{\epsfxsize=9.0cm\epsffile{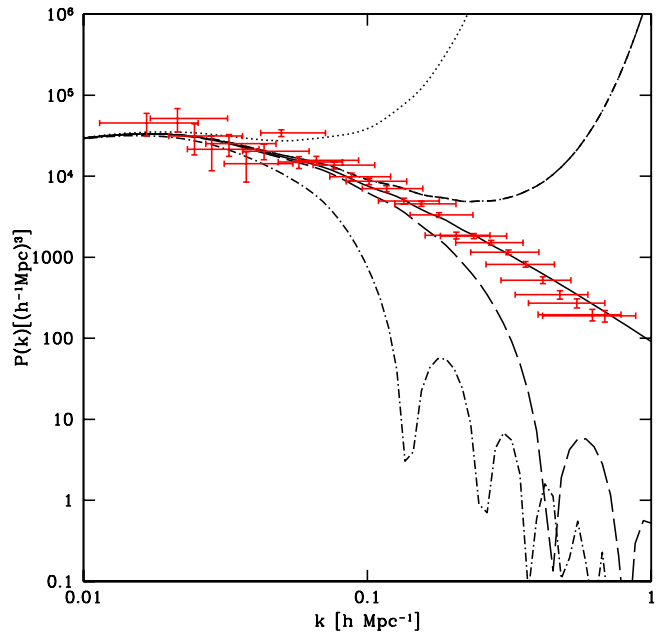}}
\caption{\label{fig1}\footnotesize UDM solution for perturbations
as function of wavenumber, $k$. From top to bottom, the 
curves are GCG models with $\alpha=-10^{-4}$, $-10^{-5}$, $0$ ($\Lambda$CDM), 
$10^{-5}$ and $10^{-4}$, respectively. 
The data points are the power spectrum of the 2df galaxy redshift survey.}
\end{figure}

For our numerical calculations, we evolved a scale invariant 
Harrison-Zeldovich spectrum up to redshift
$z=100$ (before which the GCG is indistinguishable from $\Lambda$CDM)
with CMBfast \cite{cmbfast} to correctly include
all the relevant effects (early super-horizon evolution, 
pre-recombination acoustic oscillations, Silk-damping, \etc.), with cosmological 
parameters given by the concordance model of \cite{efstathiou}.
We then used \eq{firstorder} to evolve the fluctuations from $z=100$ until today.
Results for a sample of $\alpha$-values are plotted in \fig{fig1}, and show
how tiny non-zero values of $\alpha$ result in large changes on
small scales as expected.

\begin{figure}
\centerline{\epsfxsize=9.0cm\epsffile{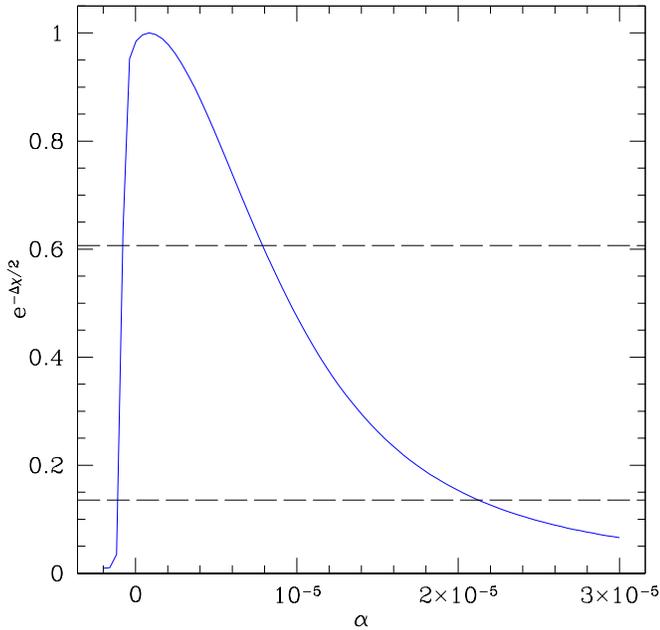}}
\caption{
The likelihood function $e^{-\Delta\chi/2}$ as a function of the GCG parameter $\alpha$. 
It is sharply
peaked around $\alpha=0$ which is equivalent to the $\Lambda$CDM
model. From top to bottom, the 
horizontal dashed lines correspond to $\Delta\chi^2=1$ and 4, respectively.
% REPETITION FROM MAIN TEXT:
%The observed skewness is due to the fact that the oscillating
%solution ($\alpha>0$)is easier to fit than the exponentially
%unstable solution ($\alpha<0$)
}
\label{likelihood}.
\end{figure}

We constrain $\alpha$ by making a $\chi^2$ fit of the theoretically predicted power spectrum
against that observed with the 2dF 100k Galaxy Redshift Survey\cite{colless} 
as analyzed by \cite{2dFGRS}.
For each $\alpha$, we use the best fitting power normalization to ensure that 
our constraints come only from the shape of the power spectrum, not from the overall 
amplitude which involves mass-to-light bias. 
To be conservative and stay firmly in the linear regime, 
we discard data with $k>0.3h/$Mpc.
We run our code for a fine grid of models with
$-1<\alpha<1$ to find the corresponding $\chi^2$ values. The
likelihood function $e^{-\Delta \chi^2/2}$ is plotted in
\fig{likelihood}. It predictably peaks around
$\alpha\approx 0$, and the observed skewness is simply due to the
fact that the oscillating solution ($\alpha>0$)is easier to fit than 
the exponentially unstable solution ($\alpha<0$).
Setting $\Delta\chi^2=1$ cutoff as in a crude Bayesian analysis
gives the constraints $-0.00000081<\alpha<0.0000079$.

\begin{figure}
\centerline{\epsfxsize=8.5cm\epsffile{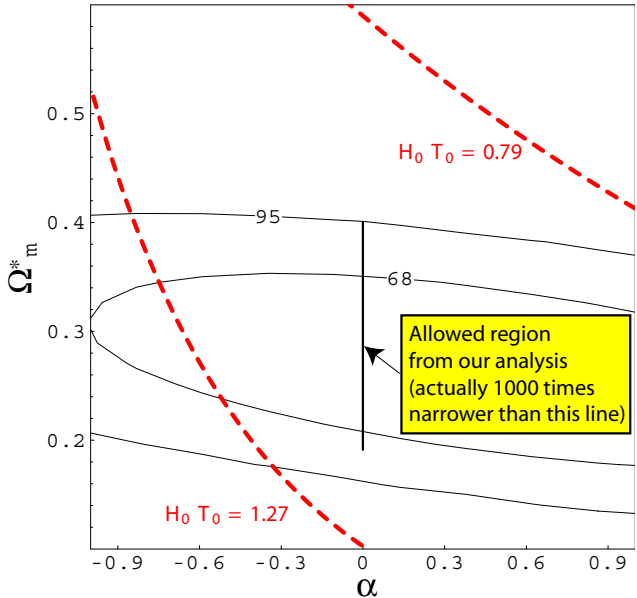}}
\caption{The graph is showing  constraints from
previous work by Makler {\protect\etal}. Our new constraints from first order
perturbation theory are superimposed on the plot as shown}
\label{exclusion}
\end{figure}

To place this result in context, \fig{exclusion} shows the 
0th order constraints from Makler {\etal} \cite{makler} 
with our new strict constraints superimposed.

\section{Conclusions}

Above we showed that GCG models with $|\alpha|\gg 10^{-5}$ are ruled out by 
observation, since they cause fluctuations or blowup in the matter power
spectrum that are not observed. Let us now examine the assumptions 
that went into this and the broader implications.

First of all, our extremely tight constraints imply that
that the narrow range of allowed GCG models will be completely indistinguishable 
from $\Lambda$CDM to both to 0th order and at the early times 
when primary CMB anisotropies are produced. This means that the corresponding
standard constraints on cosmological parameters from CMB, SN Ia {\etc} apply also 
to the GCG models making the identification $\Oms=\Om$, so that there are no 
interesting degeneracies between $\alpha$ and other parameters that can significantly
widen the allowed $\alpha$-range.
We have therefore used standard constraints $0.2<\Om<0.4$ for 
the allowed region in \fig{exclusion}. 

Second, our limiting our constraints to linear scales $k<0.3h/\Mpc$ was probably 
overly conservative. As reviewed in \cite{spacetime,pwindows}, there are quite 
strong constraints on the linear power spectrum on much smaller scales from
weak lensing, from the Lyman $\alpha$ forest and perhaps even from
lensing of halo substructure \cite{DalalPRL} which if used would tighten our
upper limit on $\alpha$ to $10^{-6}$, $10^{-7}$ and $10^{-10}$, respectively.
%lensing: 1e-5*(0.3/0.7)**2 ~ 1.8e-6
%lyaf:    1e-5*(0.3/5)**2   ~ 3.6e-8  (but at higher z so weaker constraints) 
%dalal:   1e-5*(0.3/100)**2 ~ 9e-11

Third, we saw that all that really mattered in \eq{firstorder}
as far as the constraints were concerned was 
the pressure term $(\lambdac k)^2$. This means that our results apply
more generally than merely to the GCG case: {\it any} unified dark matter
model where $p$ is a unique function of $\rho$ is ruled out
if the sound speed is non-negligible, \ie, if the function
$p(\rho)$ departs substantially from a constant over the range where pressure
has an effect --- quantitatively, if $|d\ln p/d\ln\rho|\simgt 10^{-5}$,
again rendering it indistinguishable from standard $\Lambda$CDM.

In contrast, standard quintessence models have no such problems. Although they typically have
high sound speeds causing oscillations as above, 
this does not prevent the dark matter from clustering since it is a separate dynamic component.
Quintessence models would fail as above if there the two components were tightly coupled,
and this is effectively what happens with UDM since the two are one and the same substance.

To salvage the UDM idea in some form, its pressure must not be uniquely 
determined by its density --- not even on subhorizon scales.
As worked out in detail by Hu \cite{HuGDM}, the 
effective sound speed can under some circumstances differ from the adiabatic
sound speed, and it is only the former that must approximately vanish to satisfy
our constraints.
Although it may be possible to concoct such models,
say by introducing another physical field upon which $p$ depends and making it fluctuate
in such a way as to cancel the problematic pressure gradients, 
this would be giving up much of the elegance and simplicity that gave the unified 
dark matter idea its appeal, essentially substituting one extra field for another.

In conclusion, precision data is gradually allowing us to test rather than assume 
the physics underlying modern cosmology. We have taken a step in this direction by 
ruling out a broad class of so-called unified dark matter models. Our results indicate that 
dark energy is either indistinguishable from a pure cosmological constant or 
a separate component from the dark matter with a life of its own.
%if the dark energy is a dynamic substance whose density can vary in space and time, 
%making it distinguishable from a pure cosmological constant, then it must be a separate 
%component from the dark matter.

The authors wish to thank Raul Abramo for helpful comments.
This work was supported by 
NSF grants AST-0071213, AST-0134999, AST-0098606 and PHY-0116590,
NASA grants NAG5-9194 \& NAG5-11099,
MT and MZ are David and Lucile Packard Foundation fellows and 
MT is a Cottrell Scholar of Research Corporation.

\end{document}